\documentclass[12pt]{article}
\usepackage{epsfig}

\begin{document}
\title{{\bf A NEW ANALYSIS OF THE $\pi^+\pi^-$ AND $\pi^0\pi^0 $ DATA}\thanks
{Presented by 
L. Le\'sniak at the Meson 2002, Seventh International Workshop on Production,
Properties and Interaction of Mesons, Cracow, Poland, May 24-28, 2002}}

\author{R. KAMI\'NSKI, L. LE\'SNIAK AND K. RYBICKI \\ 
\small {Henryk Niewodnicza\'nski Institute of Nuclear Physics,}
 \\ \small{ PL 31-342 Krak\'ow, Poland}}
  
\maketitle

\begin{abstract}

A joint analysis of the $\pi^+\pi^-$ and $\pi^0\pi^0 $ S-wave amplitudes 
including results of three experiments has been performed. Using our 
theoretical model of the $\pi\pi$ production in which not only 
the pion exchange but also the $a_1$ meson exchange were 
taken into account we have obtained a unique solution (called "down-flat") 
for the scalar-isoscalar phase shifts. Thus a long-standing 
problem of ambiguities in the S-wave isoscalar amplitude has been finally 
resolved and the second ("up-flat") solution was eliminated. 

\end{abstract}

  
\vspace{6mm}

Analyses of experiments on the $\pi\pi$ production processes are important 
sources of information about the properties of the scalar mesons. Our knowledge
of the basic parameters for the scalar-isoscalar mesons like $f_0(500)$, 
$f_0(980)$, $f_0(1370)$ or $f_0(1500)$ is still very limited. This makes 
tests of the QCD predictions about a possible existence of the scalar 
glueballs in the \mbox{1 to 2 GeV} range to be very difficult.

The existence of ambiguities in the $\pi\pi$ S-wave isoscalar amplitude has been
a serious problem since more than thirty years. Two experiments on the 
$\pi^-p\rightarrow \pi^+\pi^- n$ reaction, both 
using the 17.2 GeV/c $\pi^-$ beam, have been performed at CERN already more
than 20 years ago \cite{grayer,becker}. 
The second experiment was done by the CERN-Cracow-Munich Collaboration on a 
polarized target. In 1997 we have shown \cite{klr} that the data obtained in 
those two high statistics
experiments could be equally well described by at 
least two different solutions (called "up--flat" and "down--flat") for the 
$\pi\pi$ S-wave phase shifts corresponding to the \mbox{isospin 0} amplitude. 
Fortunately in 2001 new results from the E852 experiment 
performed at Brookhaven (USA) on the reaction $\pi^- p \to \pi^0\pi^0n$ at 
18.3 GeV/c have been published \cite{E852}. These $\pi^0\pi^0$ data combined
with the previous $\pi^+\pi^- $ results were used to eliminate one of the 
ambiguous solutions. Below we shall briefly describe a method to achieve this 
aim using our model of the $\pi\pi$ production \cite{klr}.
A more complete description of this analysis can be found in \cite{2002}.  

First of all we have merged two low momentum transfer squared $|t|$ bins 
measured by the E852 Group into a single one which was quite close to
the $(0.005\div0.20)~$GeV$^{2}$/c$^{2}$ bin used in the experiments performed at 
CERN. The $m_{\pi\pi}$ effective mass range chosen in this joint analysis was 
between 600 and 1460 MeV. Since the E852 collaboration has provided only the   
numbers of events and not the absolute cross sections we have determined the    
normalization comparing the intensities of the strongest $D_{0}$ wave saturated
by the $f_{2}(1270)$ resonance. The data on the reaction
$\pi^+ p \rightarrow \pi^+\pi^+ n$ have been used to calculate the $\pi\pi$ 
scalar isospin amplitude.

At high energies of incoming pions and at small momentum transfers the above 
three reactions are dominated by the one pion exchange mechanism. However 
the pseudovector $a_1(1260)$ exchange, although smaller than the one pion 
exchange, is also present and cannot be neglected.
 We consider two independent $S$--wave 
transversity amplitudes $g$ and $h$ for the reaction 
$\pi^-p_{\uparrow}\rightarrow \pi^+\pi^- n$ 
which differ by a proton or neutron spin projection on the axis perpendicular 
to the production plane. By $g_0$ and $h_0$ we denote similar amplitudes 
of the reaction $\pi^-p\rightarrow \pi^0\pi^0 n$. All these 
amplitudes have two components proportional to the off-shell 
$T(\pi\pi \rightarrow \pi\pi)$ and $t(a_1 \pi \rightarrow \pi\pi)$ amplitudes.
Both $T$ and $t$ amplitudes have two terms corresponding to isospin 0 or 2
of the pion pair. 
We use the isospin symmetry to relate different production amplitudes of the
final $\pi^+\pi^-$ or $\pi^0\pi^0$ pairs. New E852 results for the S-wave
intensity $I_0={|g_0|}^2+ {|h_0|}^2$ provide us with only one additional 
constraint at
each $m_{\pi\pi}$. This information combined with the knowledge of the 
amplitudes $g$ and $h$ from \cite{klr} is not sufficient to calculate four
independent isospin amplitudes $T_0$, $T_2$ and $t_0$, $t_2$. To proceed further
 we have 
neglected the isotensor $a_1$ exchange amplitude $t_2$ which is presumably
the smallest amplitude at low momentum transfers.
 Then one can directly relate the
amplitudes $g_0$ to $g$ and $h_0$ to $h$, namely $g_0 = 0.5 C_\pi U T_2 - g$
and $h_0 = 0.5 C_\pi U^* T_2 - h$, where $C_\pi$ and U are known functions
\cite{2002}.               
Thus using the amplitudes $g$ and $h$ found in the analysis of the $\pi^+\pi^-$
data and    
knowing the amplitude $T_2$ we can calculate the $I_0$ intensity. The results 
in the critical $m_{\pi\pi}$ range below the $K\overline{K}$ 
threshold are shown in Fig. 1. One notices that it is the "down-flat" solution 
which reasonably agrees with the E852 data. On the other hand the "up-flat" 
points are too low, especially in the region of $m_{\pi\pi}=(870\div970)$~MeV. 
\begin{figure}[htbp]\centering   
\mbox{\epsfxsize 12cm\epsfysize 7.5cm\epsfbox{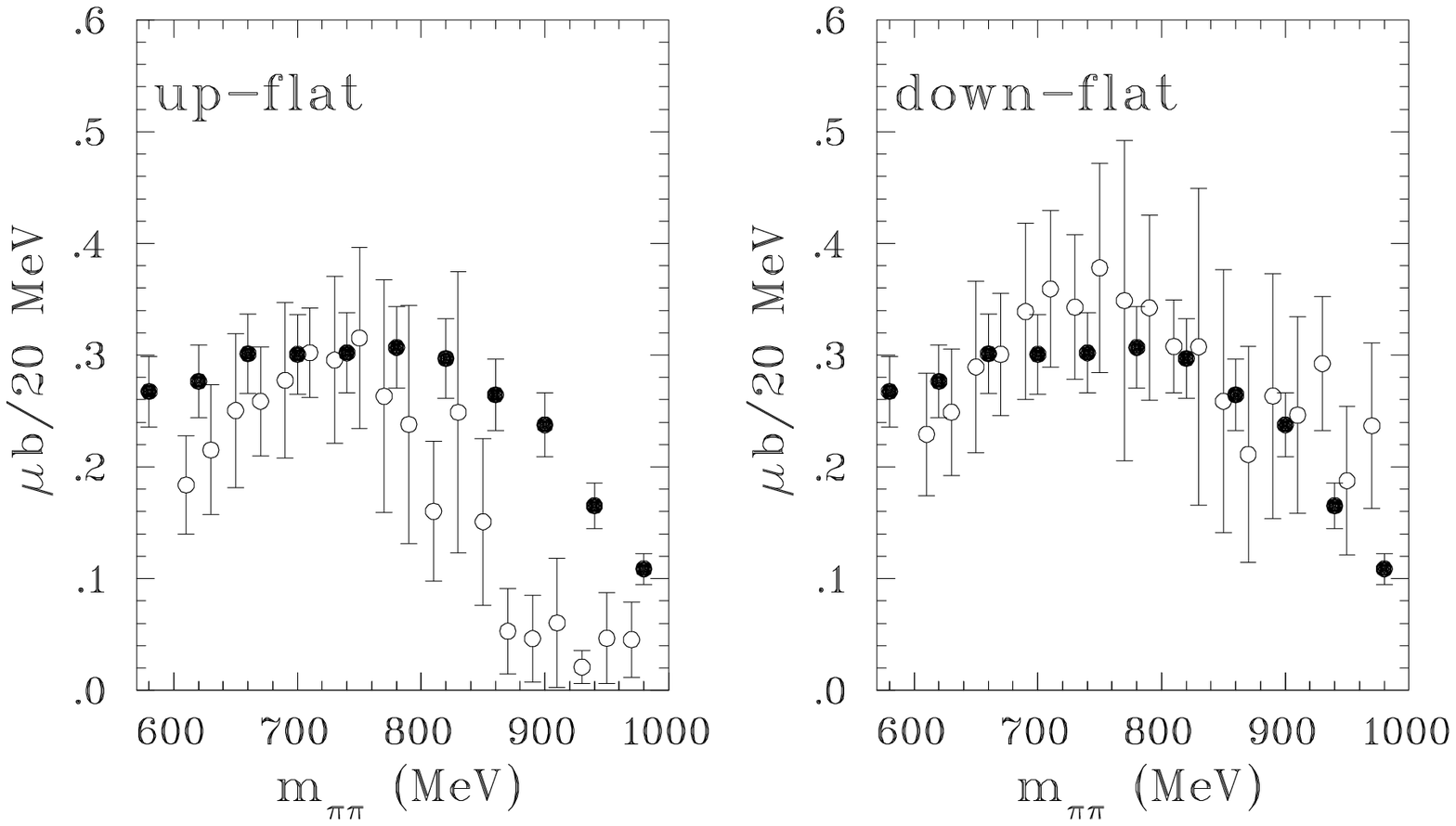}}\\   
{Fig. 1:    
A comparison of the $I_{0}$ determined from the $\pi^+\pi^-$ data (open circles)   
and the normalized S-wave intensity from the E852 $\pi^0\pi^0$ data (full 
circles)}    
\end{figure}   

 In \cite{klr} the phases   
of the transversity amplitudes $g$ and $h$ have been    
obtained essentially from the assumed resonant $\rho(770)$, $f_2(1270)$ and    
$\rho_3(1690)$ phases.
 Now we can directly calculate one phase independently for each $m_{\pi\pi}$
using the new $\pi^0\pi^0$ results for $I_0$ in addition to the $\pi^+\pi^-$
data. In this way one can adjust the phases of  $g$ and $h$ to fit 
simultaneously the $\pi^+\pi^-$ and the $\pi^0\pi^0$ cross sections. 
Then the $\pi\pi$ scalar-isoscalar phase shifts and inelasticities can be 
recalculated.  
 
Again for the "down-flat" solution to the $\pi^+\pi^-$ data we have obtained a    
reasonable behaviour of the inelasticity parameter which, as one should expect,
 is close to unity below the $K\overline{K}$ threshold.   
However, the inelasticity corresponding to the "up-flat" solution strongly 
deviates from unity. Also the "up-flat" phase shifts have an 
unphysical minimum at 900 MeV.   
This is a clear signal that the "up-flat" solution should be rejected.   
Since the "down-flat" phase shifts and inelasticities behave regularly we    
accept the "down-flat" solution, resulting from the joint analysis of the 
$\pi^+\pi^-$ and $\pi^0\pi^0 $data as a unique physical one. 


\end{document}